\documentclass{PoS}

\usepackage{amsmath}
\usepackage{dsfont}

\newcommand{\bp}{\boldsymbol{p}}

\title{Overlap Quark Propagator in Coulomb Gauge QCD}

\ShortTitle{Overlap Quark Propagator in Coulomb Gauge QCD}

\author{Ydalia Delgado Mercado\\
       Institut f\"ur Physik, Karl-Franzens Universit\"at Graz, 8010 Graz, Austria\\
        E-mail: \email{ydalia.d@uni-graz.at}}

\author{\speaker{Markus Pak}\\
        Institut f\"ur Physik, Karl-Franzens Universit\"at Graz, 8010 Graz, Austria\\
        E-mail: \email{markus.pak@uni-graz.at}}
       
\author{Mario Schr\"ock\\
       Istituto Nazionale di Fisica Nucleare (INFN), Sezione di Roma Tre, Rome, Italy\\
        E-mail: \email{mario.schroeck@roma3.infn.it}}

\abstract{The chirally symmetric Overlap quark propagator is explored in Coulomb gauge. 
This gauge is well suited for studying the relation between confinement and chiral symmetry breaking,
 since confinement can be attributed to the infrared divergent Lorentz-vector dressing function.
Using quenched gauge field configurations on a $20^4$ lattice, the quark propagator dressing functions are evaluated,
the dynamical quark mass is extracted and the chiral limit of these quantities is discussed. By removing the low-lying modes 
of the Dirac operator, chiral symmetry is artificially restored. Its effect on the dressing functions
is discussed. 
}

\FullConference{The 32nd International Symposium on Lattice Field Theory\\
		 23-28 June, 2014\\
		 Columbia University New York, NY}

\begin{document}

\section{Introduction}
The investigation of propagators is an interesting field of research, because
they can help to get a clearer picture of the non-perturbative aspects of QCD. For instance, from the Gribov-type gluon propagator
in Coulomb gauge it can be motivated why a single gluon is absent from the physical spectrum. Furthermore, analyzing the quark propagator
can yield a better understanding of the interplay between chiral symmetry breaking and confinement.
A relation between the infrared divergence of the quark dressing functions and confinement has been found in Coulomb gauge
long ago, see Ref.~\cite{Adler:1984ri, Alkofer:1988tc}, and a first step to confirm this conjecture in full QCD has been put forward
in a recent lattice study, Ref.~\cite{Burgio:2012ph}. This relation makes Coulomb gauge appealing for our purpose. 
We want to explore the confinement properties of quarks 
when chiral symmetry is artificially restored by removing the condensate from the vacuum. In recent years 
the effect of such an artificial symmetry restoration on the hadron spectrum has been analyzed, Refs.~\cite{Lang:2011vw, Glozman:2012fj}, and a
study of the Landau gauge quark propagator has been given in Ref.~\cite{Schrock:2011hq}, with using a fermion discretization, which only approximately 
fulfills the Ginsparg-Wilson equation. A chiral symmetric lattice Dirac operator is central for our purpose. 
We use Overlap fermions, Ref.~\cite{Neuberger:1997fp}, 
which offer a clear and unambiguous way to extract the dressing functions of the propagator. Studies in Landau
gauge can be found, for instance, in Refs.~\cite{Bonnet:2002ih,
Zhang:2003faa}. 

Moreover, the quark
sector in continuum Coulomb gauge is not yet satisfactory well understood, although some new insights have been established in recent years within
the so-called variational approach, see Refs.~\cite{Pak:2011wu}, \cite{Pak:2013uba}. 
Therefore a careful analysis of the dressing functions on the lattice is needed, which could then be used as input into the continuum 
bound state equations.

In this contribution we present our first results for the Overlap quark propagator in Coulomb gauge using quenched 
L\"uscher-Weisz gauge field configurations on a $20^4$ lattice. We also briefly discuss the effect of artificial 
chiral symmetry restoration on the quark propagator. The final results will be presented elsewhere, Ref.~\cite{Pak-prep}. 


\section{Lattice Setup}
\label{latt-setup}
We use quenched L\"uscher-Weisz gauge field configurations on a $20^4$ lattice with $\beta=7.552$, corresponding to 
the lattice spacing $a=0.2$ fm, Ref.~\cite{Gattringer:2001jf}. The configurations are generated using the Chroma software 
package \cite{Edwards:2004sx} and QDP-JIT \cite{Winter:2014npa, Winter:2014dka}. The average plaquette is 0.5767. 

The continuum Coulomb gauge condition $\partial_i A_i =0 \, (i=1,2,3)$ is implemented on the 
lattice via maximization of the functional 
\begin{equation}
F_g[U] = \Re\sum_{ i, x} \mathrm{tr}\big[ g(x)\, U_i(x) \, g(x+\hat{i} \,)^\dagger  \big]
\label{eq:F}
\end{equation}
with respect to gauge transformations $g(x)\in\mathrm{SU}(3)$. 
The cuLGT code, Ref.~\cite{Schrock:2012fj}, and the overrelaxation algorithm, Ref.~\cite{Mandula:1990vs} are used 
for this purpose. 

Our ensemble consists of 96 configurations with six current quark masses chosen to be in the range m = $(85-173)$ MeV. 
The Overlap operator is inverted on point-sources for each configuration. The propagators
are Fourier transformed to momentum space and the dressing functions are evaluated according to the procedure 
presented in Refs.~\cite{Burgio:2012ph, Skullerud:1999gv}. The residual gauge freedom with respect to space-independent
gauge transformations is fixed by the Integrated Polyakov gauge, Ref.~\cite{Burgio:2008jr}.

\section{Overlap fermions}
\label{overlap-chapter}
The Overlap Dirac operator  with mass parameter $m_0$ is given as
\begin{align}
  D(m_0) = \left( 1 - \frac{m_0}{2\rho} \right) D(0) + m_0 \; , \qquad D(0) = \rho \left( \mathds{1} + \gamma_5 \mbox{sign}\left[H_{\textsc{W}}(-\rho)\right] \right) \; ,
\end{align}
with the Hermitian Wilson-Dirac kernel $H_{\textsc{W}}$ and the negative Wilson mass $\rho$. We choose $\rho=1.6$ throughout this work. 
With this definition the eigenvalues lie on a circle in the complex plane with radius $\rho$. The Ginsparg-Wilson
equation is $\{D(0), \gamma_5\}=\frac{1}{\rho} D(0) \gamma_5 D(0)$.
From the free propagator in momentum space it can easily be seen that the redefinition ($S$ denotes the propagator)
\begin{align}
 \widetilde{S} = S - \frac{1}{2\rho} 
\end{align}
leads to the continuum chiral symmetry condition $\{\widetilde{S},\gamma_5 \} = 0$ at tree-level, Ref.~\cite{Bonnet:2002ih}. 
From the free massive (inverse) quark propagator
\begin{align}
 \left(\widetilde{S}^{(0)}\right)^{-1}(p) = i \gamma_{\mu} q_{\mu} + \mathds{1} m 
\end{align}
we identify the lattice momenta $q_{\mu}$ and current quark mass $m$ to be
\begin{align}
\label{qmu}
 q_{\mu} = \frac{4\rho^2}{(2\rho -m_0)}\frac{k_{\mu}\left(\sqrt{k^2_{\mu}+A^2}+A \right)}{k_{\mu}^2} \; , \quad m = \frac{m_0}{1-\frac{m_0}{2\rho}} \; ,
\end{align}
with the momenta $k_{\mu} = \sin(p_{\mu} a), \, \hat{k}_{\mu} = 2 \sin(p_{\mu} a/2) $ and $A = \frac{1}{2} \hat{k}^2_{\mu} - a \rho $. 

\section{Quark Propagator in Coulomb Gauge}
\label{propagator-chapter}
\subsection{Definition and Quark Dispersion Relation}
In Coulomb gauge the quark propagator is decomposed into four irreducible Lorentz-tensor components ($\vec{p}$ denotes three-momentum and  $p = | \vec{p}|$)
\begin{align}
\label{prop-parameterization}
 S^{-1}(p, p_4) = i \gamma_i p_i A_{\textsc{s}}(p) + i \gamma_4 p_4 A_{\textsc{t}}(p)  
+ \gamma_4 p_4 \gamma_i p_i A_{\textsc{d}}(p) + \mathds{1} B(p) \; , 
\end{align}
with $A_{\textsc{s}}, A_{\textsc{t}}, A_{\textsc{d}}, B $ referring to \textit{spatial}, \textit{temporal}, \textit{mixed} and \textit{scalar} dressing functions, respectively.
Moreover, we define the dynamical mass function $M(p)= B(p)/A_{\textsc{s}}(p)$, which gives the constituent quark mass in the 
limit $p \rightarrow 0$. The results obtained in Ref.~\cite{Burgio:2012ph} as well as our results show that all dressing functions on the
right-hand side of Eq.~(\ref{prop-parameterization}) are independent of $p_4$. It is then possible to evaluate the static quark propagator according to 
$S(p)=\int \frac{dp_4}{2\pi} S(p,p_4)$, yielding
\begin{align}
 S(p) = \frac{B(p) - i \boldsymbol{\gamma} \cdot \bp  A_{\textsc{S}}(p)}{2 \omega(p)} . 
\end{align}
The quark dispersion relation $\omega(p)$ is identified as
\begin{align}
\label{disp2}
 \omega(p) = A_{\textsc{T}}(p) A_{\textsc{S}}(p)\sqrt{p^2 + M^2(p)} \; .
\end{align}
From mean-field studies in continuum Coulomb gauge it is known that $M(p)$ approaches a constant for $p \rightarrow 0$, 
but $A_{\textsc{S}}(p)$ diverges, hence $\omega(p)$ diverges. This is how quark confinement reflects itself in the Coulomb
gauge description: via an infrared divergent energy dispersion relation $ \omega(p)$. 
In Ref.~\cite{Burgio:2012ph} a first indication of an IR-divergent dispersion relation is observed. 

\subsection{Numerical Results}
\begin{figure}[t]
 \centering  
 \includegraphics[angle=0,width=.49\linewidth]{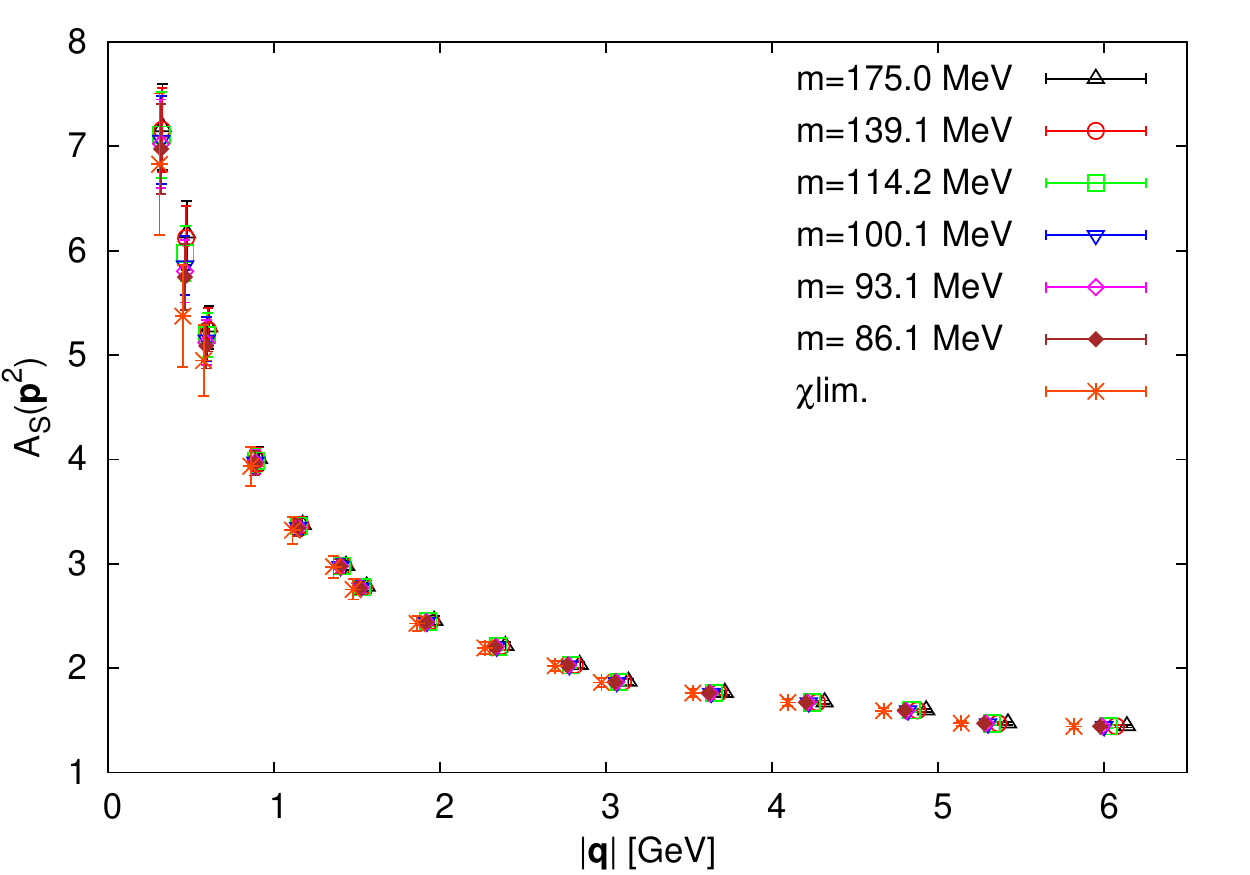}
  \includegraphics[angle=0,width=.49\linewidth]{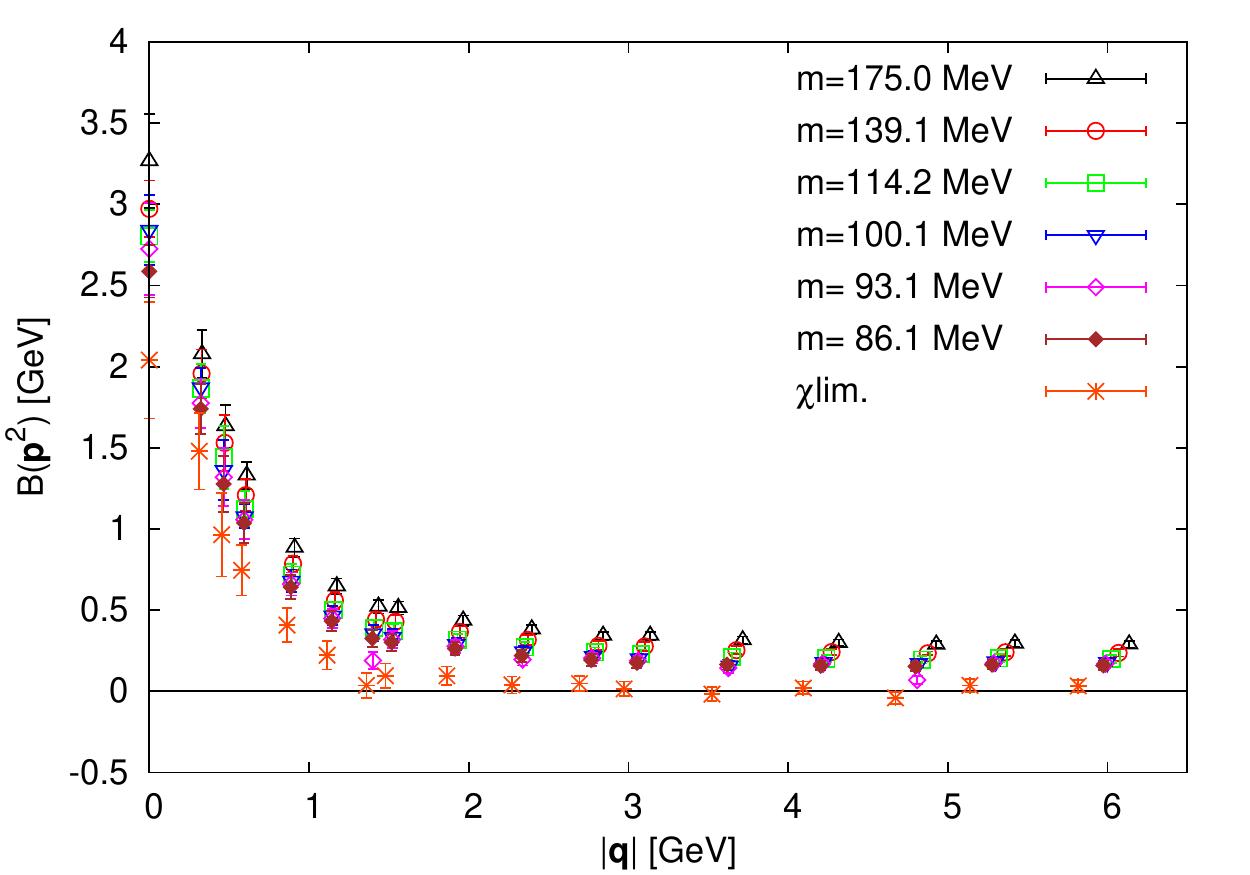}
 \caption[Spatial component for full case]{\sl
  Spatial component $A_{\textsc{s}}(\bp)$ (l.h.s.) and  scalar component $B(\bp)$ (r.h.s) as functions of 
  the lattice momentum $q$ for several quark masses and in the chiral limit.}
 \label{spatial-component}
 \end{figure}

Most interesting are the spatial and scalar dressing functions,  $A_{\textsc{s}}(p)$ and $B(p)$, see Fig.~\ref{spatial-component}. 
Due to asymptotic freedom for large momenta the vector dressing function $A_{\textsc{s}}(p)$ approaches unity. 
The scalar part $B(p)$ goes to the current quark mass and to zero in the chiral limit. 
Around 1 GeV it acquires non-zero values in the chiral limit, consistent with
chiral symmetry breaking and dynamical mass generation. 
The  scalar part $B(p)$ shows a clear dependence on the current quark mass $m$ whereas for the spatial dressing function $A_{\textsc{s}}(\bp)$ 
only a mild mass dependence is observed. 
For zero momentum, $B(p)$ reaches a finite value due to the finite volume of the lattice. 
The vector dressing function $A_{\textsc{s}}(p)$ is not accessible at zero momentum. 
We note, that in order to confirm the divergence of both dressing functions for $p\rightarrow 0$ proposed by continuum Coulomb gauge studies, 
the behavior of the dressing function towards the continuum limit has to be analyzed. This is
left to a future study. 

\begin{figure}[t]
 \centering
 \includegraphics[angle=0,width=.55\linewidth]{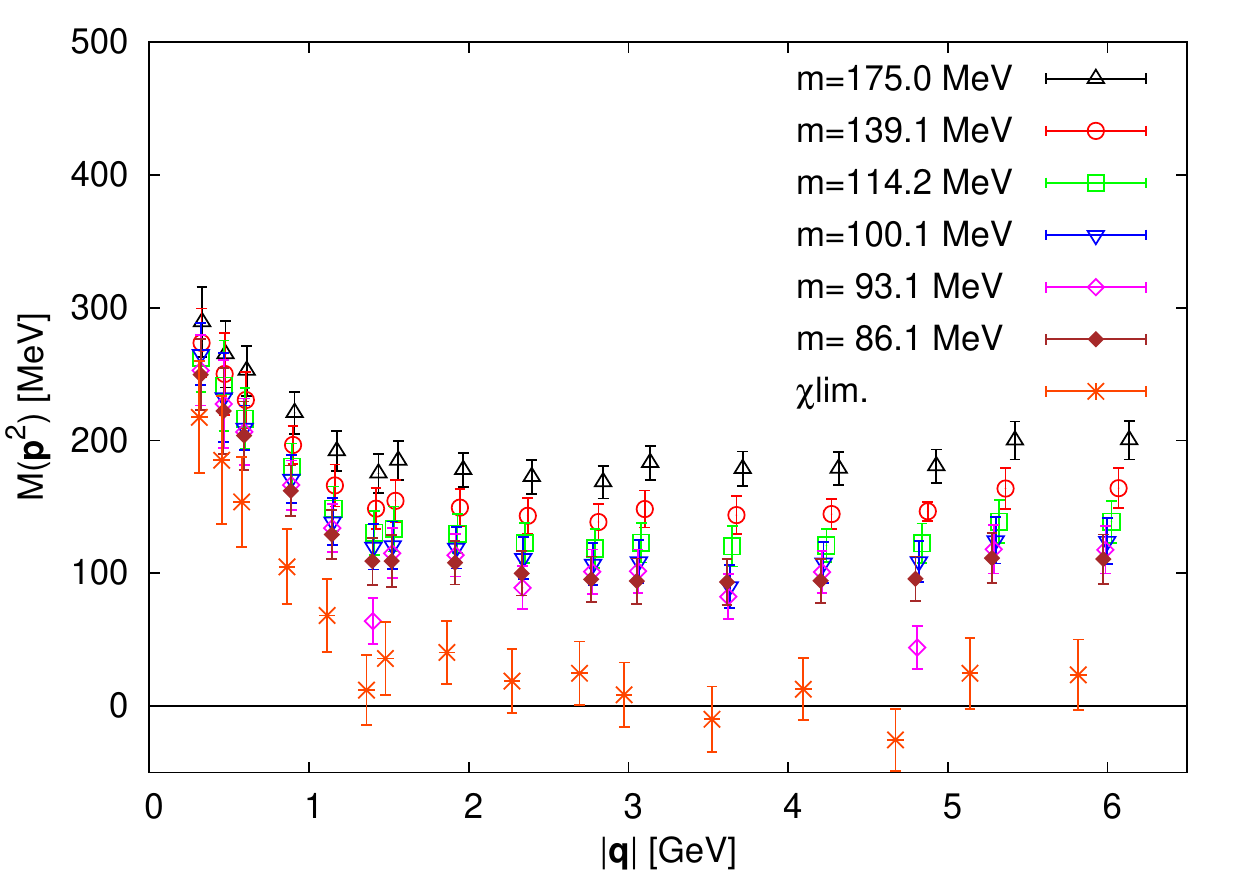}
 \caption[Lattice momenta and mass]{\sl
  Dynamical mass function $M(\bp)$ for several quark masses and in the chiral limit.}
 \label{dynamical-mass-plot}
 \end{figure}

The dynamical quark mass function $M(p)$ is shown in Fig.~\ref{dynamical-mass-plot}. For large momenta 
it approaches the current quark mass $m$ and in the chiral limit it goes to zero, like the scalar dressing function
$B(p)$. For momenta around 1 GeV 
$B(p)$ starts to increase and tends to a finite value in the IR limit, identified as constituent quark mass. 
Due to the relatively large lattice we reach IR lattice momenta $q$ around 300 MeV. 
At that value the dynamical quark mass 
is already around 200 MeV for chiral quarks. By performing a simple linear fit of the small momentum points to zero momentum, 
it is seen that a constituent quark mass around $300$ MeV is reached, which goes in hand with phenomenological predictions. 
Furthermore, it can be seen that for smaller current quark masses more dynamical mass is generated.
The most amount of dynamical mass is generated for chiral quarks. 
We also note that the dynamical quark mass $M(p)$ should only be affected by vacuum fermion loops
at the percentage level, Ref.~\cite{Burgio:2012ph}. Hence,  
we do not expect much difference for the constituent quark mass when using dynamical configurations. 

\begin{figure}[t]
 \centering
  \includegraphics[angle=0,width=.49\linewidth]{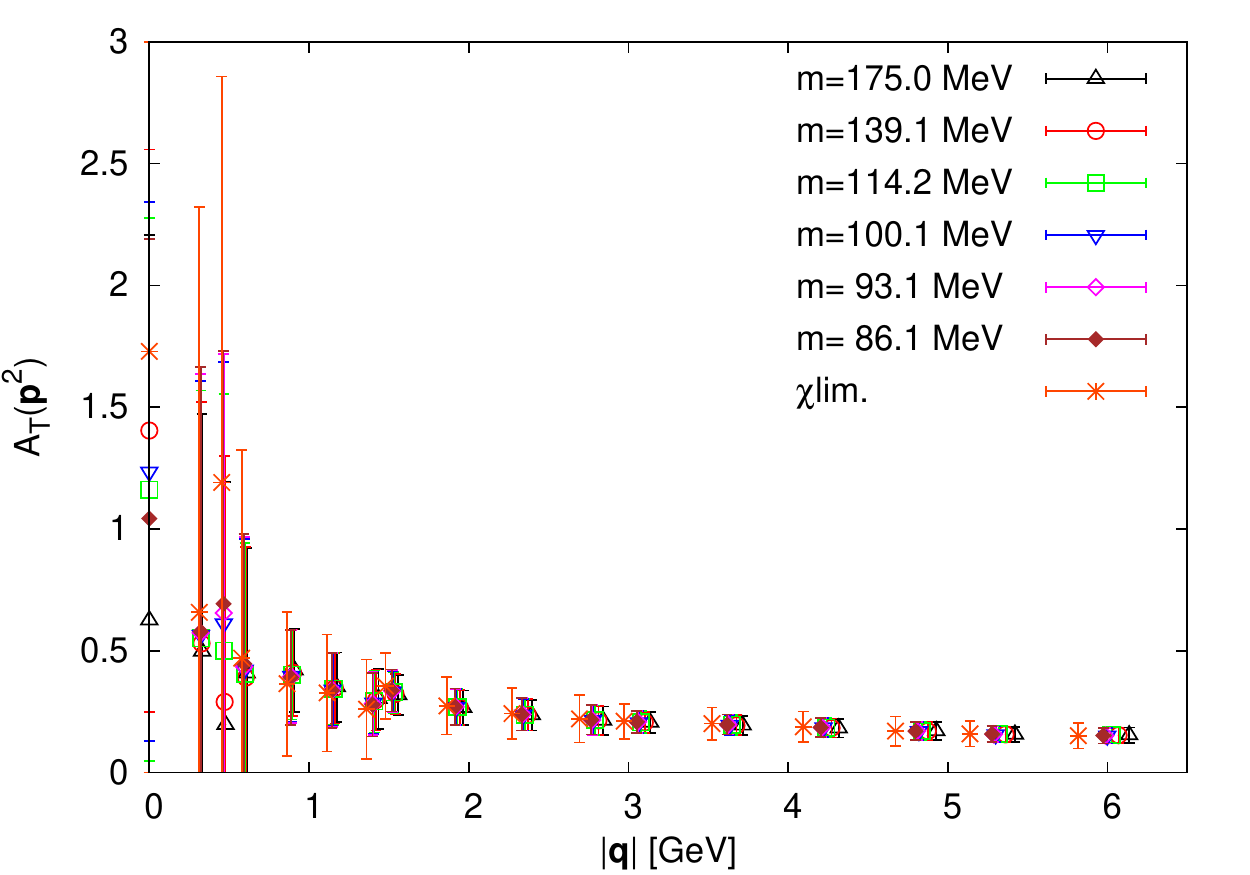}
 \includegraphics[angle=0,width=.49\linewidth]{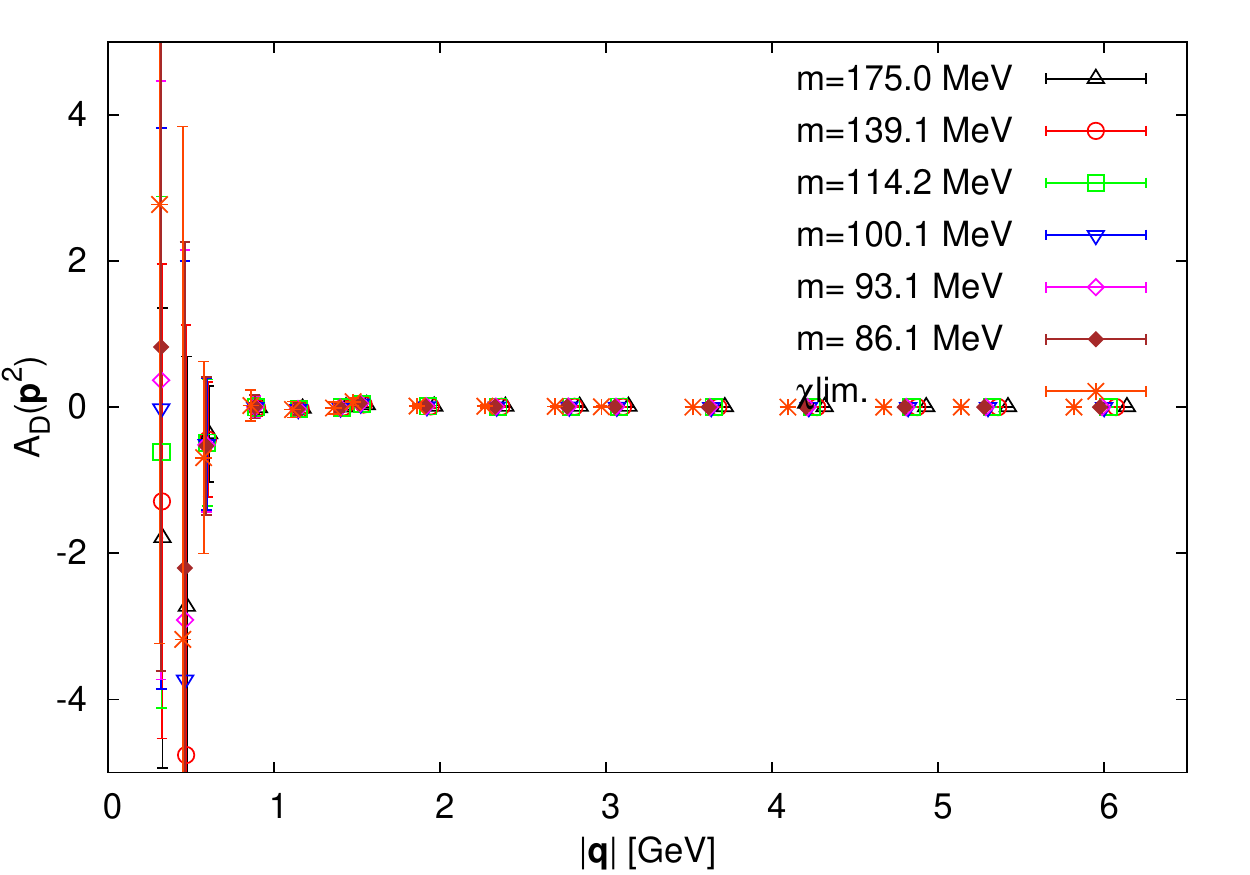}
 \caption[Temporal component for full case II]{\sl
Temporal component $A_{\textsc{t}}(\bp)$ (l.h.s.) and mixed component $A_{\textsc{d}}(\bp)$ (r.h.s.) for several current quark masses and in the chiral limit.}
 \label{mixed-component}
 \end{figure}

The temporal part $A_{\textsc{t}}(p)$ vanishes if the additional gauge freedom with respect to space independent gauge transformations
is not fixed. As in Ref.~\cite{Burgio:2012ph} we fix it to the Integrated Polyakov Gauge. In this gauge $A_{\textsc{t}}(\bp)$ goes to small non-zero values 
for large momenta, see left-hand side of Fig.~\ref{mixed-component}. For small momenta the error bars are too large to make a 
precise statement. However, a divergent behavior, as for the
dressing functions fixed by Coulomb gauge, does not seem to occur. It is more likely that a finite value is reached. 
A possible mixed component $A_{\textsc{d}}$, which does not appear at tree-level, also
seems to vanish non-perturbatively, see right-hand side of Fig~\ref{mixed-component}. Again, in the infrared region 
the error bars are too large to make a precise statement. 

\subsection{Dirac Low-Mode Removal}
When chiral symmetry is artificially restored in the vacuum, the quark condensate $\langle \overline{\psi} \psi \rangle$,
the dynamical quark mass $M(p)$ and scalar dressing function $B(p)$ vanish in the chiral limit.
The interesting question is, how the vector dressing function $A_{\textsc{s}}(p)$ is affected by chiral symmetry restoration. 
We argue that if it still increases for small momenta,
then it is likely that the dispersion relation $\omega(p)$, Eq.~(\ref{disp2}), is still divergent for $p \rightarrow 0$. However,
if it drops off, then such a divergence is not possible. We note, that the simple
mean-field Dyson-Schwinger equations suggest that $A_{\textsc{s}}(p)$ is still divergent for setting $M(p)=0$, Ref.~\cite{Glozman:2007tv}. If this picture
can be transported to full QCD is currently under investigation. First results show that the shape of $A_{\textsc{s}}(p)$ is indeed
not significantly different when removing the condensate and $A_{\textsc{s}}(p)$ still increases for small momenta. This would give a clear picture of confinement
after unbreaking of chiral symmetry in Coulomb gauge: the Lorentz-vector dressing function $A_{\textsc{s}}(p)$ survives chiral symmetry restoration
and is still infrared divergent. The final results on this issue will be presented elsewhere \cite{Pak-prep}.

\section{Summary and Outlook}
\label{summary-chapter}
Coulomb gauge offers a playground to study the interplay between confinement and chiral symmetry breaking
due to the relation between the infrared divergence of the quark propagator dressing functions and the confinement 
properties of the theory. We have shown that on a
$20^4$ lattice using quenched gauge field configurations the dressing functions increase for small momenta and 
that dynamical mass generation sets in around 1 GeV, resulting in a constituent mass for chiral quarks around $300$ MeV.
The next step is to extract the dressing functions when chiral symmetry breaking is artificially removed by
subtracting the low-mode part from the full quark propagator. For chiral quarks the dynamical mass and scalar dressing function 
approach zero in such a phase. However, first results show that the vector dressing function
stays intact, which would give a clear picture of confinement after unbreaking of chiral symmetry in Coulomb gauge. 
Then, to get a better understanding of the divergence structure of the dressing functions, the continuum limit should
be explored. In addition, Coulomb gauge could also be suitable to study the effect of instantons on the 
quark propagator, as done in Landau gauge \cite{Trewartha:2013qga}, or to study the chiral symmetry properties of adjoint quarks. 

\acknowledgments
Discussions with G.~Burgio and L.~Glozman are greatly acknowledged. 
M.P. acknowledges support by the Austrian Science Fund (FWF)
through the grant P26627-N27.  The calculations have been performed on clusters 
at ZID at the University of Graz and 
at the Graz University of Technology.


\begin{thebibliography}{99}

\bibitem{Adler:1984ri} 
  S.~L.~Adler and A.~C.~Davis,
  Nucl.\ Phys.\ B {\bf 244}, 469 (1984).

\bibitem{Alkofer:1988tc} 
  R.~Alkofer and P.~A.~Amundsen,
  Nucl.\ Phys.\ B {\bf 306}, 305 (1988).
  
\bibitem{Burgio:2012ph} 
  G.~Burgio, M.~Schr\"ock, H.~Reinhardt and M.~Quandt,
  Phys.\ Rev.\ D {\bf 86}, 014506 (2012)
  [arXiv:1204.0716 [hep-lat]].
  
\bibitem{Lang:2011vw} 
  C.~B.~Lang and M.~Schr\"ock,
  Phys.\ Rev.\ D {\bf 84}, 087704 (2011)
  [arXiv:1107.5195 [hep-lat]].

\bibitem{Glozman:2012fj} 
  L.~Y.~Glozman, C.~B.~Lang and M.~Schr\"ock,
  Phys.\ Rev.\ D {\bf 86}, 014507 (2012)
  [arXiv:1205.4887 [hep-lat]].
     
\bibitem{Schrock:2011hq} 
  M.~Schr\"ock,
  Phys.\ Lett.\ B {\bf 711}, 217 (2012)
  [arXiv:1112.5107 [hep-lat]].
  
\bibitem{Neuberger:1997fp} 
  H.~Neuberger,
  Phys.\ Lett.\ B {\bf 417}, 141 (1998)
  [hep-lat/9707022].
  
\bibitem{Bonnet:2002ih} 
  F.~D.~R.~Bonnet {\it et al.}  [CSSM Lattice Collaboration],
  Phys.\ Rev.\ D {\bf 65}, 114503 (2002)
  [hep-lat/0202003].
  
\bibitem{Zhang:2003faa} 
  J.~B.~Zhang {\it et al.}  [CSSM Lattice Collaboration],
  Phys.\ Rev.\ D {\bf 70}, 034505 (2004)
  [hep-lat/0301018].

\bibitem{Pak:2011wu} 
  M.~Pak and H.~Reinhardt,
  Phys.\ Lett.\ B {\bf 707}, 566 (2012)
  [arXiv:1107.5263 [hep-ph]].
 
\bibitem{Pak:2013uba} 
  M.~Pak and H.~Reinhardt,
  Phys.\ Rev.\ D {\bf 88}, 125021 (2013)
  [arXiv:1310.1797 [hep-ph]].
  
  
\bibitem{Pak-prep} 
  Y.~Delgado Mercado, M.~Pak and M.~Schr\"ock,
 in preparation.
  
\bibitem{Gattringer:2001jf} 
  C.~Gattringer, R.~Hoffmann and S.~Schaefer,
  Phys.\ Rev.\ D {\bf 65}, 094503 (2002)
  [hep-lat/0112024].
  
\bibitem{Edwards:2004sx} 
  R.~G.~Edwards {\it et al.}  [SciDAC and LHPC and UKQCD Collaborations],
  Nucl.\ Phys.\ Proc.\ Suppl.\  {\bf 140}, 832 (2005)
  [hep-lat/0409003].

\bibitem{Winter:2014npa} 
  F.~Winter,
  PoS LATTICE {\bf 2013}, 042 (2013).

\bibitem{Winter:2014dka} 
  F.~T.~Winter, M.~A.~Clark, R.~G.~Edwards and B.~Jo\'{o},
  arXiv:1408.5925 [hep-lat].
  
\bibitem{Schrock:2012fj} 
  M.~Schr\"ock and H.~Vogt,
  Comput.\ Phys.\ Commun.\  {\bf 184}, 1907 (2013)
  [arXiv:1212.5221 [hep-lat]].
  
\bibitem{Mandula:1990vs} 
  J.~E.~Mandula and M.~Ogilvie,
  Phys.\ Lett.\ B {\bf 248}, 156 (1990).
  
\bibitem{Skullerud:1999gv} 
  J.~I.~Skullerud and A.~G.~Williams,
  Nucl.\ Phys.\ Proc.\ Suppl.\  {\bf 83}, 209 (2000)
  [hep-lat/9909142].
  
\bibitem{Burgio:2008jr} 
  G.~Burgio, M.~Quandt and H.~Reinhardt,
  Phys.\ Rev.\ Lett.\  {\bf 102}, 032002 (2009)
  [arXiv:0807.3291 [hep-lat]].

\bibitem{Glozman:2007tv} 
  L.~Y.~Glozman and R.~F.~Wagenbrunn,
  Phys.\ Rev.\ D {\bf 77}, 054027 (2008)
  [arXiv:0709.3080 [hep-ph]].
  
\bibitem{Trewartha:2013qga} 
  D.~Trewartha, W.~Kamleh, D.~Leinweber and D.~S.~Roberts,
  Phys.\ Rev.\ D {\bf 88}, 034501 (2013)
  [arXiv:1306.3283 [hep-lat]].


\end{thebibliography}
\end{document}